\documentclass[sigconf,authorversion=True]{acmart}

\usepackage{natbib}
\usepackage{amsmath, amsfonts}
\usepackage{url}
\usepackage{graphicx}
\usepackage{color}
\usepackage{booktabs} 
\usepackage{siunitx, color}
\usepackage{multirow}
\usepackage{algorithm, algpseudocode}
\usepackage{caption, subcaption}

\renewcommand\vec{\mathbf}
\algnewcommand\algorithmicinput{\textbf{Input:}}
\algnewcommand\INPUT{\item[\algorithmicinput]}
\algnewcommand\algorithmicoutput{\textbf{Output:}}
\algnewcommand\OUTPUT{\item[\algorithmicoutput]}

\begin{document}
\title{A Hybrid Approach to Music Playlist Continuation Based on Playlist-Song Membership}


\author{Andreu Vall, Matthias Dorfer, Markus Schedl, Gerhard Widmer}
\affiliation{
  \institution{Johannes Kepler University, Department of Computational Perception}
}
\email{andreu.vall@jku.at}

\renewcommand{\shortauthors}{A. Vall et al.}

\copyrightyear{2018}
\acmYear{2018}
\setcopyright{acmlicensed}
\acmConference[SAC 2018]{SAC 2018: Symposium on Applied Computing}{April 9--13, 2018}{Pau, France}
\acmBooktitle{SAC 2018: Symposium on Applied Computing, April 9--13, 2018, Pau, France}
\acmPrice{15.00}
\acmDOI{10.1145/3167132.3167280}
\acmISBN{978-1-4503-5191-1/18/04}

\begin{abstract}
Automated music playlist continuation is a common task of music recommender systems, that generally consists in providing a fitting extension to a given playlist. Collaborative filtering models, that extract abstract patterns from curated music playlists, tend to provide better playlist continuations than content-based approaches. However, pure collaborative filtering models have at least one of the following limitations: (1) they can only extend playlists profiled at training time; (2) they misrepresent songs that occur in very few playlists. We introduce a novel hybrid playlist continuation model based on what we name ``playlist-song membership,'' that is, whether a given playlist and a given song fit together. The proposed model regards any playlist-song pair exclusively in terms of feature vectors. In light of this information, and after having been trained on a collection of labeled playlist-song pairs, the proposed model decides whether a playlist-song pair fits together or not. Experimental results on two datasets of curated music playlists show that the proposed playlist continuation model compares to a state-of-the-art collaborative filtering model in the ideal situation of extending playlists profiled at training time and where songs occurred frequently in training playlists. In contrast to the collaborative filtering model, and as a result of its general understanding of the playlist-song pairs in terms of feature vectors, the proposed model is additionally able to (1)~extend non-profiled playlists and (2)~recommend songs that occurred seldom or never in training~playlists.
\end{abstract}
%

%
%



\maketitle

\section{Introduction}
\label{sac2018_membership:sec:introduction}

The automated continuation of music playlists enables music recommendation scenarios such as playing and sequentially extending a music stream (similar to traditional radio broadcasting) or suggesting to the user fitting songs to extend their own music playlists. In both cases, it is crucial to identify candidate songs that fit a given playlist, and this is a particularly challenging question. By analyzing interviews with practitioners and postings to a dedicated playlist-sharing website, \citet{cunningham_more_2006} identified that the playlist curation process is complex and influenced by a variety of factors like mood, theme or purpose. Furthermore, they found that the agreement among practitioners on curation rules was only reduced to rather loose guidelines.

A successful approach to music playlist continuation relies on mining collaborative information through the analysis of curated music playlists. In particular, collaborative approaches based on statistical models explain curated playlists in terms of a defined quantitative criterion, providing a principled approach to modeling the fitness of songs in playlists. In contrast to pure content-based approaches, collaborative approaches tend to reveal more abstract patterns over playlists and songs, but the good performance of collaborative approaches is strongly dependent on the availability of a large volume of training data. This requirement is easily compromised. Firstly, the amount of carefully curated playlists is rather scarce (especially compared to the abundant--but not curated--listening logs derived from music streaming services). Secondly, music consumption is inescapably biased towards popular songs~\cite{celma_music_2010}, resulting in a vast majority of songs occurring in very few playlists.

We introduce a novel hybrid playlist continuation model that combines curated music playlists with multimodal song features. The curated music playlists are used to derive training examples of playlist-song pairs that fit and playlist-song pairs that do not fit. The multimodal song features make the proposed model robust to data scarcity problems. The proposed model is designed to flexibly evaluate the fitness of any playlist-song pair, making it possible to extend any playlist by selecting suitable songs among any set of song candidates. In contrast to previous hybrid playlist continuation models, the proposed model is a \emph{feature-combination} hybrid~\cite{burke_hybrid_2002}, having the advantage that the collaborative information and the song features are implicitly fused into a single enhanced recommender system.

The remainder of this paper is organized as follows. Section~\ref{sac2018_membership:sec:related_work} reviews the related work. Section~\ref{sac2018_membership:sec:model} presents the hybrid playlist continuation model. The evaluation methodology is described in Section~\ref{sac2018_membership:sec:evaluation}. Section~\ref{sac2018_membership:sec:datasets} describes the datasets of curated playlists and song features used in our experiments. Section~\ref{sac2018_membership:sec:results} elaborates on the results. Finally, conclusions are drawn in Section~\ref{sac2018_membership:sec:conclusion}.

\section{Related Work}
\label{sac2018_membership:sec:related_work}

A well-researched approach to automated music playlist continuation relies on the song content. Pairwise song similarities are computed on the basis of features extracted from the audio signal (possibly enriched with social tags and metadata) and used to enforce content-wise smooth transitions~\cite{logan_content-based_2002,mcfee_natural_2011, pohle_generating_2005,knees_combining_2006, flexer_playlist_2008}. Recommendations based on content similarity are expected to confer coherence to the playlist. However, pure content-based recommendations can not capture complex relations and, in fact, it does not hold in general that the songs in a playlist should all sound \mbox{similar~\cite{lee_how_2011}.}

Collaborative Filtering (CF) has been proven successful to reveal underlying structure from, in general, user-item interactions~\cite{ricci_recommender_2015, adomavicius_toward_2005}. In particular, CF has been applied to music playlist continuation by regarding each playlist as a user's listening history, on the basis of which songs should be recommended. Previous research has mostly focused on playlist-neighborhood CF models~\cite{hariri_context-aware_2012,bonnin_automated_2014,jannach_beyond_2015}, but \citet{aizenberg_build_2012} also present a latent-factor CF model tailored to mine Internet radio stations, accounting for song artist, time of the day and song adjacency. An important limitation of most latent-factor and playlist-neighborhood CF models is that they need to profile the playlists at training time in order to extend them, by computing their latent factors or finding their nearest neighbors. As a consequence, such models can not extend playlists unseen at training time. To circumvent this issue, \citet{aizenberg_build_2012} replace the latent factors of unseen playlists by the latent factors of their songs, and \citet{jannach_when_2017} show how to efficiently implement a playlist-neighborhood CF model that can extend unseen playlists in reasonable time. Song-neighborhood CF models can in general extend unseen playlists,  because they only require pairwise song similarities~\cite{sarwar_item-based_2001,vall_importance_2017}. A common limitation of all pure CF methods is that they are only aware of the songs occurring in the training playlists. Thus, songs that never occurred in the training playlists, to which we refer as ``out-of-set'' songs, can not be recommended.

\citet{zheleva_statistical_2010} propose a latent-variable playlist model based on Latent Dirichlet Allocation (LDA)~\cite{blei_latent_2003}. They found that a variation of the basic LDA model that takes listening sessions into consideration provides better playlist continuations. In a similar line, \mbox{\citet{chen_playlist_2012}} present a playlist model named Latent Markov Embedding, that exploits radio playlists to learn an embedding that projects songs into a Euclidean latent space. Both models can extend playlists without precomputing playlist profiles, but they can only recommend songs occurring in training playlists.

Hybrid models combining CF and song features are a common approach to mitigate the difficulties of CF models to represent infrequent songs. \citet{hariri_context-aware_2012} represent the songs in hand-curated playlists by topic models derived from social tags and then mine frequent sequential patterns at the topic level. The scores of a CF model are re-ranked according to the next topics predicted. The approach proposed by \mbox{\citet{jannach_beyond_2015}} pre-selects suitable next songs based on the weighted combination of scores yielded by a playlist-neighborhood CF model and content-based similarities. The song candidates are then re-ranked to match the recent songs. In both cases, the hybridization follows from the combination of independently obtained scores by means of weighting heuristics or re-ranking. For example, the CF prediction for a song occurring only in a few training playlists would be boosted with content-based information. However, the prediction for an out-of-set song would solely rely on the content-based component.

Similar to our approach, \citet{van_den_oord_deep_2013} also relate song content with collaborative patterns using a deep neural network. The network is trained to predict the CF factors of a song given its log-compressed mel-spectrogram. However, the two approaches are fundamentally different. Our approach integrates collaborative information and song features into a standalone enhanced recommendation model that can decide if any playlist-song pair matches. In contrast, the model proposed in \cite{van_den_oord_deep_2013} tries to transform audio features into more abstract collaborative features, which still need to be processed to yield a final recommendation. Also, its performance is naturally upper-bounded by CF.

For a more comprehensive survey on music playlist continuation, we point the interested reader to \cite{bonnin_automated_2014} or Chapter 13 in \cite{ricci_recommender_2015}.

\section{Hybrid Playlist Continuation}
\label{sac2018_membership:sec:model}


In this section we introduce the proposed hybrid playlist continuation model, including: basic concepts and notation, the data used to train it, the model definition, how to use it to recommend playlist continuations and finally some implementation details.

\subsection{Basic Concepts and Notation}
\label{sac2018_membership:sec:notation}

Let $P$ be a collection of music playlists. Let $S$ be the universe of songs available, including at least all the unique songs occurring in the playlists of $P$, but possibly more. A song $s \in S$ is represented by a feature vector {$\vec{x}_s \in \mathbb{R}^D$}, where $D$ is the song-feature dimensionality. A playlist $p \in P$ of length $T_p$ is regarded as a set of songs and it is represented by a feature matrix $\vec{X}_p \in \mathbb{R}^{T_p \times D}$ that contains, in each row, the feature vector of each song in the playlist. Different playlists may have different lengths. 

We want to remark that, by considering a playlist as a set of songs, we are disregarding its song order. This assumption may seem counterintuitive because the process of listening to a playlist is inherently sequential. However, our preliminary studies on the importance of song order in curated music playlists indicate that the order is actually not crucial to recommend continuations to such playlists. Even though more research is required to fully understand the impact of the song order, we feel confident that disregarding the song order does not harm the contribution of the current work.

Throughout the paper, we take advantage of standard set operations and notation~\cite[chap.1]{kaplansky_set_1972} to precisely express relations between playlists and songs. We advance here common expressions that we normally use. A playlist $p$ is a set of songs and thus it is a subset of the universe of songs, i.e., $p \subseteq S$. Then, the set difference $S \setminus p$ corresponds to all the songs in the universe $S$ that do not belong to the playlist~$p$. A song $s$ can be regarded as playlist of one song, i.e., as the singleton set $\{s\}$. Given a song in a playlist $s \in p$, the set difference $p \setminus \{s\}$ removes the song $s$ from the playlist~$p$. Despite the use of set operations and notation, for the sake of clarity we continue using the word ``length'' instead of ``cardinality'' to refer to the number of songs in a playlist.

\subsection{Playlist-Song Training Examples}
\label{sac2018_membership:sec:training_examples}

The proposed hybrid playlist continuation model is based on what we name ``playlist-song membership,'' that is, whether a given playlist and a given song fit together. In Section~\ref{sac2018_membership:sec:related_work} we have discussed the complexity of deciding which songs fit together and how approaches that exploit collaborative information are able to reveal more abstract relations than content-based approaches. Therefore we use curated playlists as a form of collaborative implicit feedback~\cite{hu_collaborative_2008, pan_one-class_2008} to derive training examples of playlist-song pairs that fit, as well as of playlist-song pairs that do not fit.

To ease the reading we say that a playlist-song pair that fits is a ``match,'' while a playlist-song pair that does not fit is a ``mismatch.'' In the context of implicit feedback it might be more accurate to use the term ``no-match'' rather than ``mismatch,'' to stress the fact that missing feedback does not necessarily reflect negative feedback, but we keep the latter for simplicity.

Our basic assumption is that any playlist $p \in P$ implicitly defines matches with its own songs, in the sense that any song $s \in p$ matches the shortened playlist $p_s = p \setminus s$, i.e., the original playlist $p$ where the song $s$ has been removed. We further assume that any song not occurring in the playlist $p$ is a mismatch to the shortened playlist $p_s$. Thus a mismatch is obtained by randomly drawing a song from the remaining songs~$S \setminus p$. In this way we can derive training examples of matching and mismatching playlist-song pairs. Precisely, we follow Algorithm~\ref{sac2018_membership:algo:assemble_train_set}, that given a playlists collection $P$ and a universe of songs $S$ yields as many matching as mismatching playlist-song pairs.
\begin{algorithm}
\caption{Derive playlist-song matches and mismatches.}
\label{sac2018_membership:algo:assemble_train_set}
\begin{algorithmic}[1]
	\INPUT
		\Statex $P$ \Comment{playlists collection}
		\Statex $S$ \Comment{universe of songs}
	\OUTPUT
		\Statex matches \Comment{list of playlist-song matches}
		\Statex mismatches \Comment{list of playlist-song mismatches}
	\Statex
	\State matches $=$ \texttt{[]} \Comment{initialize empty lists}
	\State mismatches $=$ \texttt{[]}	
	\For{$p \in P$}
		\For{$s \in p$}
			\State $p_s = p \setminus \{s\}$  \Comment{remove $s$ from $p$}
			\State $s_+ = s$ \Comment{$s$ is a match to $p_s$}
			\State $s_- = \texttt{sample}(S \setminus p)$  \Comment{draw a mismatch to $p_s$}
			\State matches.\texttt{append}$\left( (p_s, s_+) \right)$ \Comment{store train. examples}
			\State mismatches.\texttt{append}$\left( (p_s, s_-) \right)$
		\EndFor
	\EndFor
	\State \Return matches, mismatches
\end{algorithmic}
\end{algorithm}

It is important to observe that, in general, considering playlist-song pairs observed in curated playlists as matches, and any playlist-song pairs not observed in curated playlists as mismatches, would yield many more mismatches than matches. This strong class imbalance is well-known in the domain of recommender systems~\cite{adomavicius_toward_2005,ricci_recommender_2015} and can be tackled with different approaches such as cost-sensitive learning~\cite{pan_one-class_2008,hu_collaborative_2008} or sampling techniques~\cite{maimon_data_2010}. As we have seen in Algorithm~\ref{sac2018_membership:algo:assemble_train_set}, we opt for the latter, and derive a balanced number of matching and mismatching training examples.

The training examples derived from Algorithm~\ref{sac2018_membership:algo:assemble_train_set} are formatted in order to use them within the proposed playlist continuation model. Each playlist-song pair $(p, s)$ is represented by its feature matrix and vector $(\vec{X}_p, \vec{x}_s)$, and it is labeled with $y_{p, s} \in \{0, 1\}$, where 1 indicates a matching playlist-song pair and 0 indicates a mismatching playlist-song pair. The final training dataset consists of all the triplets $\{(\mathbf{X}_p, \mathbf{x}_s), y_{p,s}\}$.

\subsection{Model Definition and Learning}
\label{sac2018_membership:sec:model_definition}

\begin{figure}
 \centerline{
 \includegraphics[scale=0.7]{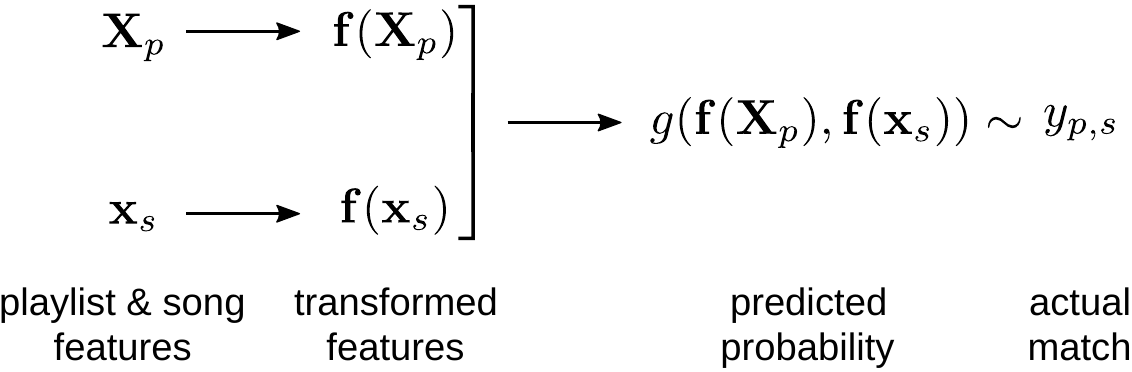}
 }
 \caption{\small Sketch of the hybrid playlist continuation model. Given any playlist-song pair, its feature matrix and vector are transformed to hidden representations that are then used to decide if the playlist-song pair is a match or a mismatch. The model is trained on labeled playlist-song pairs derived from Algortithm~\ref{sac2018_membership:algo:assemble_train_set}.}
\label{sac2018_membership:fig:network_sketch}
\end{figure}

The proposed hybrid playlist continuation model has to be able to decide if any given playlist-song pair constitutes a match or a mismatch. We generally devise this model as a deep neural network consisting of a ``feature-transformation'' component $\vec{f}$ and a ``match-discrimination'' component $g$~(Figure~\ref{sac2018_membership:fig:network_sketch}). The feature-transformation component takes any playlist-song pair $(p, s)$ as input, represented by the corresponding feature matrix and vector $(\vec{X}_p, \vec{x}_s)$. The song feature vector $\vec{x}_s$ is transformed to an \mbox{$H$-dimensional} hidden representation $\vec{f}(\vec{x}_s) \in \mathbb{R}^H$.  The playlist feature matrix $\vec{X}_p$ is song-wise subject to the same transformation, obtaining a hidden representation \mbox{$\vec{f}(\vec{X}_p) \in \mathbb{R}^{T_p \times H}$} (where we slightly abuse notation for~$\vec{f}$). Both hidden representations are passed together through the match-discrimination component that predicts the probability of the playlist-song pair being a match: $g\left( \vec{f}(\vec{X}_p), \vec{f}(\vec{x}_s) \right) \in [0, 1]$.

The transformation component $\vec{f}$ and the match-discrimination component $g$ depend on sets of learnable weights $\boldsymbol{\theta}_\vec{f}$ and $\boldsymbol{\theta}_g$, respectively (omitted so far for the sake of clarity). The weights are adjusted on the basis of the training examples $\{(\mathbf{X}_p, \mathbf{x}_s), y_{p,s}\}$ derived from Algorithm~\ref{sac2018_membership:algo:assemble_train_set}, by comparing the model's predicted match probability for a pair $(p, s)$ to the actual match label \mbox{$y_{p, s} \in \{0, 1\}$}. Precisely, the sets of weights $\boldsymbol{\theta}_\vec{f}$, $\boldsymbol{\theta}_g$ are estimated to minimize the following binary cross-entropy cost function
\begin{equation}
\begin{split}
\label{sac2018_membership:eq:cost_function}
\mathcal{L} \Bigl( \boldsymbol{\theta}_{\vec{f}}, \boldsymbol{\theta}_g & \bigm| \left\{ (\vec{X}_p, \vec{x}_s), y_{p,s} \right\} \Bigr) = \\
- & \sum_{p, s} y_{p,s} \log \left( \hat{y}_{p,s} \right) + \left( 1 - y_{p,s} \right) \log \left( 1 - \hat{y}_{p,s} \right)\text{,}
\end{split}
\end{equation}
where $\hat{y}_{p,s} = g \bigl(\vec{f}(\vec{X}_p), \vec{f}(\vec{x}_s) \bigr)$ is the model's prediction for the playlist-song pair $(p, s)$.

In Section~\ref{sac2018_membership:sec:introduction} we pointed out that the proposed playlist continuation model is a feature-combination hybrid~\cite{burke_hybrid_2002}, that is, it implicitly fuses content-based and collaborative information into a joint recommender system. Equation~\ref{sac2018_membership:eq:cost_function} clearly shows how the proposed model relies on matching and mismatching playlist-song pairs derived from curated music playlists, while regarding playlist-song pairs in terms of their feature representations.

\subsection{Playlist Continuation}
\label{sac2018_membership:sec:continuation}

Once it has been trained, the proposed model is used to predict playlists continuations in the following way. Suppose that we recommend a continuation to playlist $p$ using the universe of songs~$S$. To avoid the recommendation of songs already present in the playlist, we restrict the set of candidate songs to $S \setminus p$. We let the model predict the matching probability of the playlist-song pair~$(p, s)$, for each candidate song $s \in S \setminus p$. This operation is linear in the number of candidate songs. We then rank the candidate songs in order of preference to extend $p$. The ranked list of song candidates can be used differently depending on the recommendation scenario. For example, in case of extending a radio stream one could recommend the top result, or assisting a user in identifying relevant songs to extend a playlist, one could show the top $K$ results, or the results with predicted matching probability above a threshold. We detail the evaluation methodology followed for our experiments in Section~\ref{sac2018_membership:sec:evaluation}.

\subsection{Model Implementation}
\label{sac2018_membership:sec:model_implementation}
The results presented in this work (Section~\ref{sac2018_membership:sec:results}) were obtained by using the model architecture detailed in \mbox{Table~\ref{sac2018_membership:tab:model_architecture}}. The model hyperparameters were selected on a withheld validation set, choosing those that yielded best validation cost. The model was implemented using Lasagne~\cite{dieleman_lasagne:_2015} and Theano \cite{theano_development_team_theano:_2016}.
\begin{table}
\caption{\small Architecture of the hybrid playlist continuation model. The input to the model are the features $(\vec{X}_p, \vec{x}_s)$ of a playlist-song pair $(p, s)$. $\vec{X}_p^t$ denotes the $t$-th row of the feature matrix $\vec{X}_p$, i.e., the $t$-th song of the playlist $p$. The upper part of the table corresponds to the feature transformation component $\vec{f}$, and the lower part of the table corresponds to the match-discrimination component $g$. The boldface layers DE\textsubscript{k}, BN\textsubscript{k} in the transformation component $\vec{f}$ share their weights for all the songs in the playlist $p$ and for song $s$~(Section~\ref{sac2018_membership:sec:model_definition}). The dimensionality of each layer is annotated in parentheses. DE:~Dense layer, RE:~Rectify non-linearity, BN:~Batch Normalization \cite{ioffe_batch_2015},  DR:~Dropout~\cite{srivastava_dropout:_2014}.}
\centering
\includegraphics[width=0.47\textwidth]{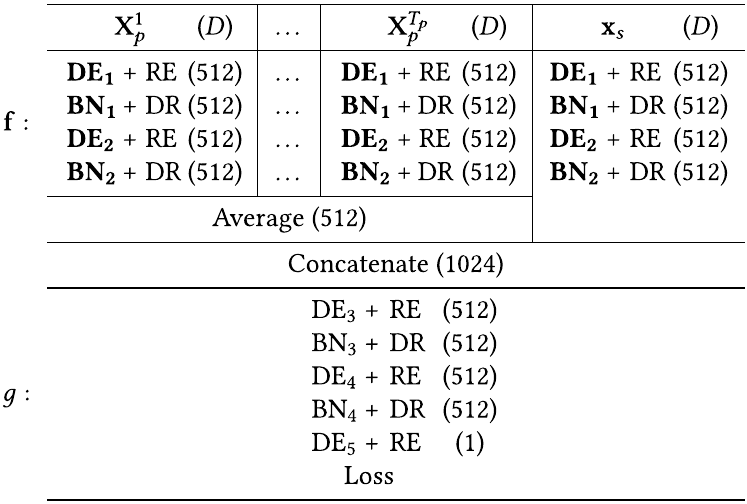}
\label{sac2018_membership:tab:model_architecture}
\end{table}

\section{Evaluation}
\label{sac2018_membership:sec:evaluation}

A large-scale on-line evaluation where users could assess the quality of the playlist continuations recommended by the proposed model should be the preferred option \cite{ricci_recommender_2015}, but would require a complex infrastructure beyond the scope of this work. Instead we design an off-line evaluation experiment, similar to previous approaches in the literature~\cite{aizenberg_build_2012,bonnin_automated_2014,hariri_context-aware_2012,jannach_beyond_2015}, where we assess the ability of the proposed model to recover withheld playlist continuations. Even though off-line experiments can not faithfully assess the quality of playlist continuations as would do it a user, we deem reasonable to assume that if a playlist continuation approach consistently achieves better off-line performance than another, it will likely perform better in practice as well.

\subsection{Off-line Experiment and Metrics}
\label{sac2018_membership:sec:experiment_metrics}

As a reminder, let $P$ be a collection of music playlists and let $S$ be the universe of songs available, including at least all the unique songs occurring in the playlists of $P$, but possibly more. Given a playlist~$p$, we assume that a continuation $p_c$, proportionally shorter than $p$, is known and withheld for test. This assumption on the length of the continuation $p_c$ follows the evaluation methodology used by \citet{aizenberg_build_2012}, but differs from the approach of \citet{bonnin_automated_2014,hariri_context-aware_2012} and \citet{jannach_beyond_2015}, where the withheld continuations have a single song regardless of the length of the playlist $p$. 

To prevent the model from recommending songs already present in the playlist $p$, we only consider the songs in $S \setminus p$, and rank them according to the model's predicted probability that they match $p$. Since this is an off-line experiment, now we do not really prepare a recommended playlist continuation for a user to evaluate. Instead, we compute different evaluation metrics based on the ability of the playlist model to rank the songs within the playlist continuation $p_c$ in top positions of the list of ranked song candidates. For each song in the withheld continuation $p_c$, we compute its rank within the whole list of ranked song candidates. For the whole continuation $p_c$, we compute the average precision given the whole list of ranked song candidates. Finally, for the whole continuation $p_c$, we compute the recall within lists of top~10, top~30 and top~100 ranked song candidates~\citep[chap.8]{manning_introduction_2009}. Even though lists of top~30 or top~100 song candidates may seem impractical for actual recommendation scenarios, these are common list lengths used to evaluate playlist continuations in off-line \mbox{experiments~\cite{aizenberg_build_2012,hariri_context-aware_2012,bonnin_automated_2014,jannach_beyond_2015}.}

This process is repeated for all the playlists we set to extend, and a summary of the described evaluation metrics over all the playlist continuations is reported, namely the median rank, the mean average precision (MAP) and the mean recall at 10, 30 and~100.

\subsection{Weak and Strong Generalization}
\label{sac2018_membership:sec:weak_strong_generalization}

\begin{figure}
 \centerline{
 \includegraphics[scale=0.27]{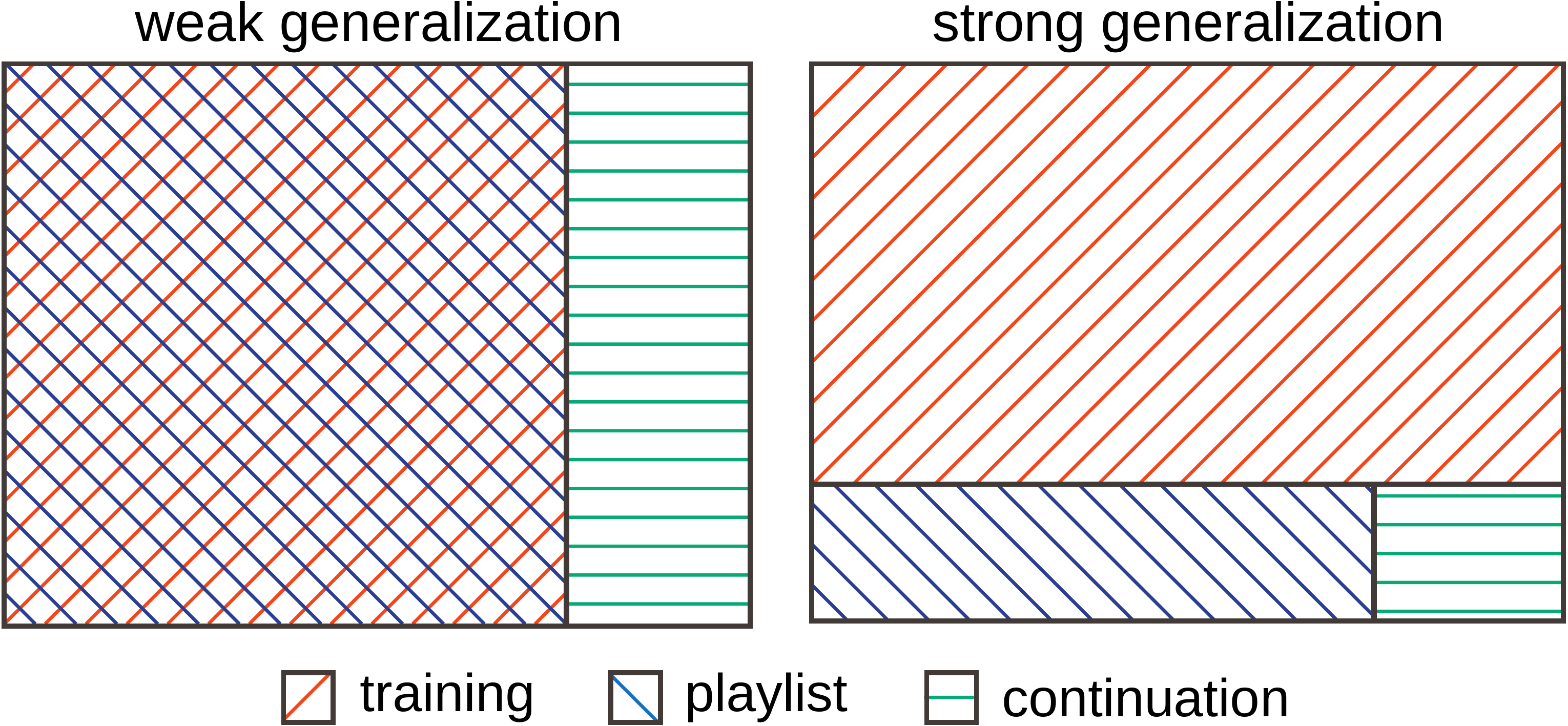}
 }
 \caption{\small Illustration of the considered evaluation settings. We figuratively picture the datasets as matrices where playlists are stored in rows, and each element in a playlist is a song. The red stripes indicate the playlists used to train the model. The blue stripes indicate the playlists that the model has to extend. The green stripes indicate the withheld continuations used for evaluation. The weak generalization setting considers a set of playlists and their withheld continuations. The model is trained on the playlists. Once trained, we assess the ability of the model to extend the very same training playlists. The strong generalization setting considers two independent sets of playlists. The first set of playlists is used to train the model. Once trained, we assess the ability of the model to extend the playlists from the second set of playlists.}
\label{sac2018_membership:fig:evaluation}
\end{figure}

We consider two different evaluation settings (Figure~\ref{sac2018_membership:fig:evaluation}), that were also proposed by~\citet{aizenberg_build_2012}. The first setting, or ``weak generalization'' setting, considers a single set of playlists with their corresponding withheld continuations. The playlists are used to train the proposed playlist continuation model. Once trained, the model predicts lists of ranked song candidates to extend the very same training playlists. The model is then evaluated using the withheld continuations as describe above. We refer to this evaluation setting  as ``weak,'' because the playlist continuation model extends playlists that it has seen before. In Section~\ref{sac2018_membership:sec:related_work} we discussed that some playlist continuation models (e.g., a latent-factor CF model), need to compute a playlist profile at training time in order to extend a playlist. Such models can operate in the weak generalization setting. The second setting, or ``strong generalization'' setting, considers two independent sets of playlists. The first set of playlists does not need known withheld continuations because it is only used to train the model, while the second set of playlists does require withheld continuations for evaluation. The playlist continuation model is trained on the playlists from the first set. Once trained, the model is shown the playlists from the second set and it predicts lists of ranked song candidates to extend them. The model is then evaluated using the withheld continuations as described in Section~\ref{sac2018_membership:sec:experiment_metrics}. We refer to this evaluation setting as ``strong,'' because the playlist continuation model extends playlists that it has never seen before. Playlist continuation models that require a precomputed playlist profile to extend a playlist can no operate in the strong generalization setting.

\subsection{Collaborative Filtering Baseline}
\label{sac2018_membership:sec:baseline}

We compare the proposed playlist continuation model to a latent-factor CF baseline. We choose the state-of-the-art Weighted Matrix Factorization (WMF) model introduced by \citet{hu_collaborative_2008} because it is specifically designed to perform CF on implicit feedback datasets. We apply WMF to the task of playlist continuation by regarding playlists as listening histories. We set a matrix with as many rows as playlists and as many songs as unique songs in the playlists. In other words, the general CF user-item matrix becomes a playlist-song matrix. If a given playlist contains a given song, then the corresponding cell in the matrix is set to 1. Otherwise, the cells are set to 0. The original model by \citet{hu_collaborative_2008} expects frequency information for each observation, e.g., the number of times a user viewed a website or the play counts of a user for a specific song. This frequency information is leveraged to assign appropriate confidence levels to each observation in the matrix. Curated playlists lack such frequency information because they are \emph{static} lists compiled by practitioners, shared on-line and finally encoded as binary values that only indicate which songs belong to the playlists. Thus, we assign the same weight to all the observations, but we tune this weight to yield best validation cost on a withheld validation set. We also experiment with different weights for the L2-regularization term. We use the publicly available implementation of WMF by~\mbox{\citet{frederickson_fast_2017}.} 

WMF can only operate on the weak generalization setting because it needs to precompute playlist profiles, in this case playlists factors, in order to extend the playlists. Furthermore, as any pure CF model, WMF can only recommend songs occurring in training playlists. Details regarding the comparison of the proposed playlist continuation model and this CF baseline are included in Section~\ref{sac2018_membership:sec:results}.

\section{Datasets}
\label{sac2018_membership:sec:datasets}

We evaluate the proposed playlist continuation model on two datasets of hand-curated playlists, along with song features derived from the Million Song Dataset\footnote{\url{https://labrosa.ee.columbia.edu/millionsong}}~(MSD)~\cite{bertin-mahieux_million_2011}.

The ``AotM-2011'' dataset~\cite{mcfee_hypergraph_2012} is a playlists collection derived from the Art of the Mix on-line database.\footnote{\url{http://www.artofthemix.org}} Each playlist in the AotM-2011 dataset is represented by a list of song titles, artist names, and links to the MSD identifiers, where available. We also have access to a private playlists collection from ``8tracks,''\footnote{\url{https://8tracks.com}} an on-line platform where users share playlists and that supports listening to them through streaming. Each playlist in the 8tracks collection is represented by song titles and artist names. We use fuzzy string matching to resolve them against the MSD, adapting the code released by \citet{jansson_this_2015} for a very similar task. The string matching results in roughly 2,5M song identifiers from the 8tracks dataset (many are spelling duplicates) resolved into 241,123 song identifiers from the MSD. Linking the 8tracks dataset to the MSD enables the extraction of song features and makes the comparison to the AotM-2011 dataset fair.

\subsection{Playlists Datasets}
\label{sac2018_membership:sec:playlists_datasets}

The AotM-2011 dataset contains a considerable number of artist- and album-themed playlists. One may argue that such playlists should be considered if they are important to playlist curators, but we deem important to exclude them for two main reasons. Firstly, we presume that playlists with several songs by the same artist or from the same album may correspond to a not so careful compilation process. Secondly, and most importantly, we utilize song features that are partly derived from social tags (Section~\ref{sac2018_membership:sec:song_features}). By manual inspection we observe that some tags inform about the artist or the album. Thus, by rejecting artist- and album-themed playlists we prevent evaluation problems regarding leaking artist or album information. Note that, in any case, excluding these playlists makes the problem harder. 
To discard artist- and album-themed playlists we keep only the playlists with at least 7 unique artists and with a maximum of 2 songs per artist (the thresholds were manually chosen to yield sufficient training playlists after the whole filtering process). The 8tracks dataset has not artist- or album-themed playlists because the terms of use of the 8tracks platform do not allow users to include in a playlist more than 2 songs by the same artist or from the same album. However, we apply the same filters to both datasets for the sake of consistency. To ensure that the model learns from playlists of sufficient length, we further filter both datasets by keeping only the playlists with at least 14 songs linked to the MSD. However, we also discard songs for which features can not be extracted due to missing raw data and therefore the final length of the playlists may be shortened.

In order to set up the training and the test sets, we discard playlists that have become shorter than 5 songs after the song filtering. For the weak generalization setting, we split each playlist leaving approximately 80\% of the songs as a training playlist and the remaining 20\% of the songs as a withheld continuation. For the strong generalization setting, we split the AotM-2011 and the 8tracks playlists collections into two playlist-disjoint subcollections each. In each case, the first playlists subcollection includes 80\% of the playlists, that will be used as training playlists. The second playlists subcollection includes 20\% of the playlists, that are further split leaving approximately 80\% of the songs as a playlist to be extended, and the remaining 20\% of the songs as a withheld continuation. Figure~\ref{sac2018_membership:fig:evaluation} illustrates the different playlists splits in the weak and strong generalization settings.


The filtered AotM-2011 dataset has 2,715 playlists with 12,355 songs by 4,097 artists. The filtered 8tracks dataset has 3,272 playlists with 14,613 songs by 5,119 artists. Detailed statistics regarding the distribution of unique songs per playlist, unique artists per playlist and song frequency in the dataset are included in Table~\ref{sac2018_membership:table:stats_playlists}.
\begin{table*}
  \centering
  \captionsetup{width=0.75\textwidth}
  \caption{\small Descriptive statistics for the playlists within the AotM-2011 and the 8tracks playlists datasets. We report the distribution of the number of songs per playlist, the number of artists per playlist, and the song frequency in the dataset (i.e., the number of playlists in which each song occurs).}
  \label{sac2018_membership:table:stats_playlists}
  \begin{tabular}{rlS[table-format=1]S[table-format=1]S[table-format=1]S[table-format=2]S[table-format=3]|S[table-format=1]S[table-format=1]S[table-format=1]S[table-format=1]S[table-format=2]}
    \toprule
    && \multicolumn{5}{c|}{Training set} & \multicolumn{5}{c}{Test set} \\
    && \text{min} & \text{1q} & \text{med} & \text{3q} & \text{max} & \text{min} & \text{1q} & \text{med} & \text{3q} & \text{max} \\
    \midrule
	\multirow{3}{*}{AotM-2011}	& Songs per playlist	& 4 & 6 & 7 & 9 & 21 & 1 & 1 & 2 & 2 & 5	\\
								& Artists per playlist	& 3	& 6 & 7 & 9 & 21 & 1 & 1 & 2 & 2 & 5	\\
								& Song frequency		& 1	& 1 & 1 & 2	& 35 & 1 & 1 & 1 & 1 & 11	\\
    \midrule
	\multirow{3}{*}{8tracks}	& Songs per playlist 	& 4 & 6 & 8 & 10 & 30  & 1 & 2 & 2 & 2 & 8	\\
								& Artists per playlist	& 3	& 6 & 8 & 10 & 28  & 1 & 2 & 2 & 2 & 8	\\
								& Song frequency		& 1	& 1 & 1 & 2	 & 119 & 1 & 1 & 1 & 1 & 27	\\
    \bottomrule
  \end{tabular}
\end{table*}

\subsection{Song Features}
\label{sac2018_membership:sec:song_features}

The proposed playlist continuation model requires each song in the playlists to be represented by a feature vector. We extract state-of-the-art song features, namely i-vectors from audio, word2vec semantic features from social tags and collaborative song latent factors from independent listening logs. We concatenate these features into a rich multimodal song feature vector. The choice of this specific set of features is motivated by our recent study~\cite{vall_music_2017}, that elaborates on the performance of these and other song features for the task of music playlist continuation, including the individual performance of each type of feature, as well as their joint performance when they are combined into a multimodal feature vector. The feature extraction process is outlined below, but we refer the interested reader to \cite{vall_music_2017} for a comprehensive presentation.

We derive the features from the MSD, together with the accompanying ``Last.fm Dataset''\footnote{\url{https://labrosa.ee.columbia.edu/millionsong/lastfm}} and the ``Taste Profile Subset.''\footnote{\url{https://labrosa.ee.columbia.edu/millionsong/tasteprofile}} The raw data available is the following. For the audio content, the MSD splits songs into segments of variable length (typically under a second) and provides 12-dimensional timbral coefficients (similar to MFCCs) for each segment. The Last.fm Dataset provides song-level social tags along with weights describing their relevance. Finally, the Taste Profile Subset includes user-song play counts derived from independent listening logs. The MSD also provides other song features such as ``danceability'' or ``energy,'' that are expected to summarize aspects of the audio at a high-level. However, these features are not documented in the MSD nor were they in their original source, the now discontinued Echo Nest API.\footnote{\url{http://the.echonest.com}} Thus we do not consider them for research purposes.

The extraction of i-vectors and word2vec semantic features requires pretraining reference models on a large collection of representative songs. For both the AotM-2011 and the 8tracks datasets, we select playlists with at least 10 songs linked to the MSD, by at least 5 artists, such that no artist has more than~2 songs in the playlist. We assume that the unique songs in the resulting playlists are representative. For each dataset, we further exclude the songs that appear only in the corresponding weak-generalization training split to minimize leaking information in the evaluation. We refer to the final song collections as the ``development song sets.'' For the AotM-2011 dataset we obtain 48,393~songs and for the 8tracks dataset we obtain 47,617~songs.

For all the feature types we extract 200-dimensional feature vectors. According to our experiments in this and our previous work~\cite{vall_music_2017}, vectors of this dimensionality carry enough information.

\subsubsection{I-Vectors from Timbral Features.}
I-vectors were first introduced in the field of speaker verification~\cite{dehak_front-end_2011}, but recently they have also been successfully utilized for music similarity and music artist recognition tasks~\cite{eghbal-zadeh_i-vectors_2015,eghbal-zadeh_timbral_2015}. We build a Gaussian mixture model with 1,024 components on the entire pool of segment-level features of the development song set. Using the unique songs in the playlists, we derive the total variability space yielding \mbox{200-dimensional} \mbox{i-vectors}. Following the standard i-vector extraction pipeline, we further transform the obatained i-vectors using a linear discriminant analysis model~\cite{hastie_elements_2008} fit on the training playlists.

\subsubsection{Semantic Features from Social Tags}
We collect the social tags of all the songs in the development song set and build a music-aware text corpus by fetching the English Wikipedia\footnote{\url{https://en.wikipedia.org}} pages of the collected tags. We run the continuous bag-of-words algorithm\footnote{\url{https://code.google.com/p/word2vec}} on the text corpus to obtain a dictionary of 200-dimensional semantic features for the most relevant words in the corpus. For each unique song in the playlists, we look up its social tags in the dictionary. If a tag is a compound of several words (e.g., ``pop rock''), we compute the average feature. Since a song may have several tags, the final semantic feature is the weighted average of all its tags' features, where the weights are provided by the Last.fm Dataset and indicate how relevant is each tag for each song.

\subsubsection{Latent Factors from Listening Logs}

We factorize the user-song play counts from the MSD using the WMF model~\cite{hu_collaborative_2008}, that is specially designed for implicit feedback datasets. We use a depth of 200 factors. We discard the user latent factors, unrelated to the playlist continuation problem, and keep only the song latent factors.

\subsubsection{Multimodal Feature Vectors}

For each song in the playlists, we concatenate its i-vector, word2vec semantic feature vector and collaborative song factors into a multimodal feature vector. Since each individual feature vector is 200-dimensional, the final multimodal feature vector is 600-dimensional.

\section{Results}
\label{sac2018_membership:sec:results}

We evaluate the proposed playlist continuation model by conducting the off-line experiment described in Section~\ref{sac2018_membership:sec:experiment_metrics}, in both the weak and the strong generalization settings proposed in Section~\ref{sac2018_membership:sec:weak_strong_generalization}, using both the AotM-2011 and the 8tracks datasets. We also evaluate the CF baseline presented in Section~\ref{sac2018_membership:sec:baseline}. Since it can only extend playlists for which playlist factors have been computed at training time, we can only evaluate it in the weak generalization setting. We finally evaluate a random baseline that given any playlist-song pair predicts a random probability of it being a match. The performance of the random baseline is independent of the generalization setting. 

Table \ref{sac2018_membership:table:results} reports the performance metrics achieved by each of the playlist continuation models. In the weak generalization setting, the proposed playlist continuation model and the CF baseline show comparable performance in both datasets. The proposed model is able to rank the songs from the withheld continuations roughly 200 positions higher (better) than the CF baseline. On the other hand, the CF baseline achieves slightly higher recall values than the proposed model. By comparison to the random baseline, we see that the proposed model and the CF baseline reveal non trivial patterns in the data. However, we also note that the absolute performance metrics are rather low. \citet{platt_learning_2002} and \mbox{\citet{mcfee_natural_2011}} also pointed out the low performances achieved when automated music playlist continuation is evaluated as an information retrieval task. They explained it by the nature of the playlist continuation problem, namely a playlist may be extended by a number of potentially relevant songs, but information retrieval off-line metrics only accept exact matches to the withheld continuations. In this regard, it is also worth noting that related works on the AotM-2011 and the 8tracks datasets report results comparable to ours~\cite{bonnin_automated_2014,jannach_beyond_2015}. We also observe that both the proposed model and the CF baseline achieve slightly worse results on the AotM-2011 dataset than in the 8tracks dataset. Since the differences are comparable for both models, we believe that this is due to the specific properties of each dataset.
\begin{table*}
  \centering
  \captionsetup{width=0.8\textwidth}  
    \caption{\small Results achieved by the proposed playlist continuation model, the CF baseline and a random baseline on the off-line evalauation experiment. We report the median rank, the MAP and the recall (R) at lists of top10, top30 and top100 song candidates. The median rank compares to 12,355 song candidates for the AotM-2011 dataset and to 14,613 song candidates for  the 8tracks dataset. Lower is better. For MAP and R@\{10, 30, 100\} higher is better.}
  \label{sac2018_membership:table:results}
  \begin{tabular}{rllS[table-format=4]S[table-format=1.2]S[table-format=1.2]S[table-format=1.2]S[table-format=1.2]}
    \toprule
    \textbf{dataset} & \textbf{generalization} & \textbf{model} & \textbf{med rank} & \textbf{MAP} & \textbf{R@10} & \textbf{R@30} & \textbf{R@100} \\
    \midrule
    AotM-2011	& weak				& proposed  	& 1230      & 1.35\%    & 2.62\%    & 6.32\%    & 13.89\%   \\
				&					& CF			& 1444      & 1.96\%    & 3.99\%    & 7.84\%    & 14.56\%   \\    
    \cmidrule(lr){2-8}
				& strong			& proposed  	& 1395      & 1.10\%	& 1.93\%    & 4.63\%	& 11.65\%   \\
				& 					& CF			& {\textemdash} & {\textemdash} & {\textemdash} & {\textemdash} & {\textemdash} \\
	\cmidrule(lr){2-8}
				& weak \& strong	& random		& 6087      & 0.11\%    & 0.20\%    & 0.28\%    & 0.79\%    \\
    \cmidrule(lr){1-8}    
    8tracks		& weak				& proposed		& 726		& 2.31\%    & 4.34\%    & 9.39\%   	& 20.02\%   \\
				&					& CF			& 1000		& 2.65\%    & 5.06\%    & 10.14\%   & 19.60\%   \\
    \cmidrule(lr){2-8}
				& strong			& proposed  	& 706      & 2.41\%		& 4.20\%    & 9.58\%    & 19.36\%   \\
				&					& CF			& {\textemdash} & {\textemdash} & {\textemdash} & {\textemdash} & {\textemdash} \\
    \cmidrule(lr){2-8}
				& weak \& strong	& random		& 7320      & 0.09\%    & 0.12\%    & 0.23\%    & 0.66\%    \\
                
    \bottomrule
  \end{tabular}
\end{table*}

As we have discussed in Section~\ref{sac2018_membership:sec:weak_strong_generalization}, the proposed playlist continuation model can evaluate any playlist-song pair, regardless of whether the playlist and the song were observed before. This flexibility enables the proposed model to extend playlists in the strong generalization setting. However, the question is whether the proposed model's performance is harmed in this evaluation setting. Still in Table~\ref{sac2018_membership:table:results}, we observe that the median rank in the strong generalization setting is roughly 150 positions lower (worse) than in the weak generalization for the AotM-2011 dataset. However, the median rank remains stable for the 8tracks dataset. Similarly, the recall values in the strong generalization setting are slightly lower in the AotM-2011 dataset, but comparable for the 8tracks dataset. Overall, even though the strong generalization setting poses a much harder task, the proposed model shows a performance comparable to its own performance in the weak generalization setting.

\begin{figure*}
\captionsetup[subfigure]{aboveskip=-2pt, belowskip=5pt}
    \centering
    \begin{subfigure}[b]{0.49\textwidth}
        \includegraphics[width=\textwidth]{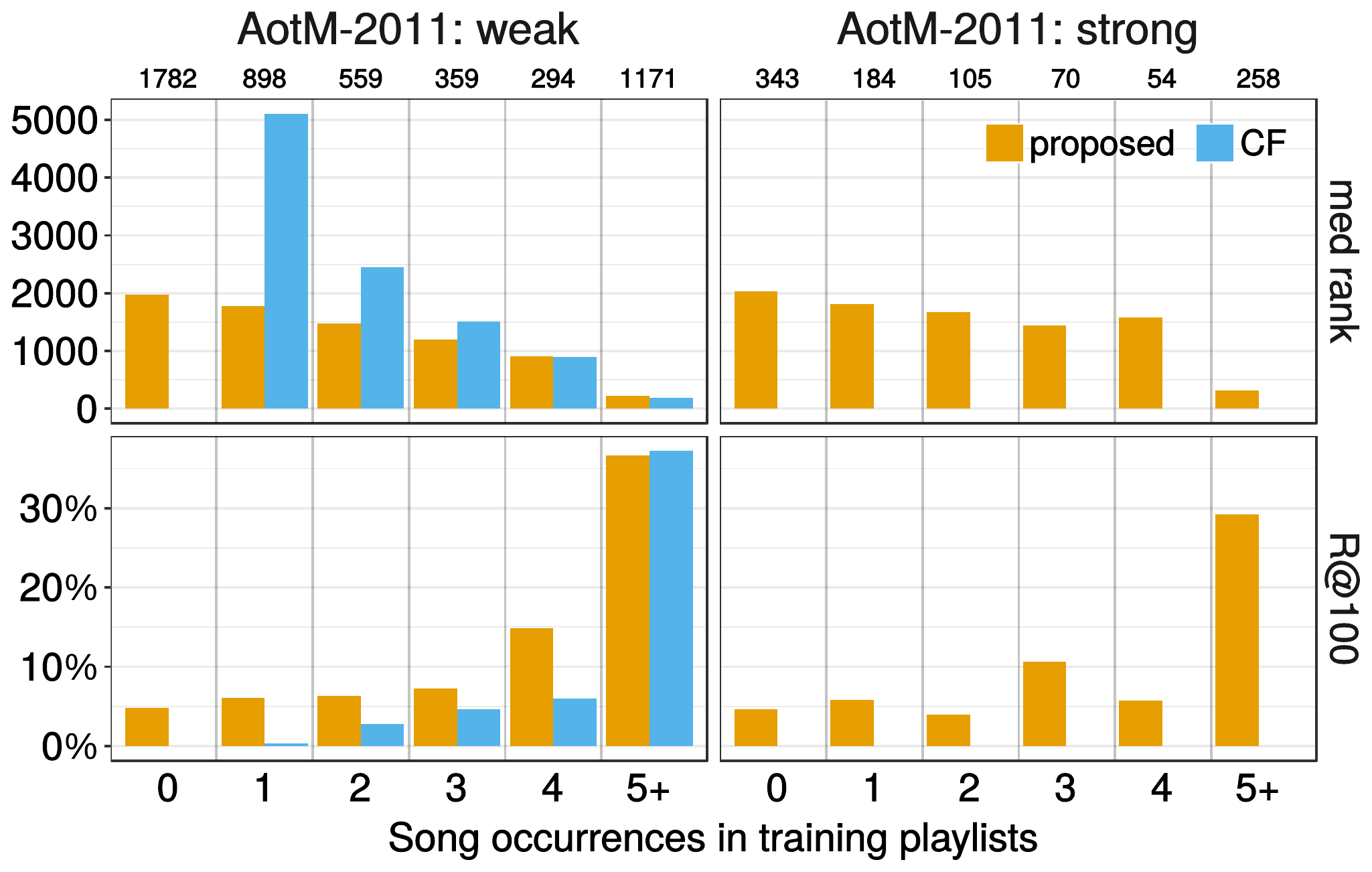}
    \end{subfigure}
    \begin{subfigure}[b]{0.49\textwidth}
        \includegraphics[width=\textwidth]{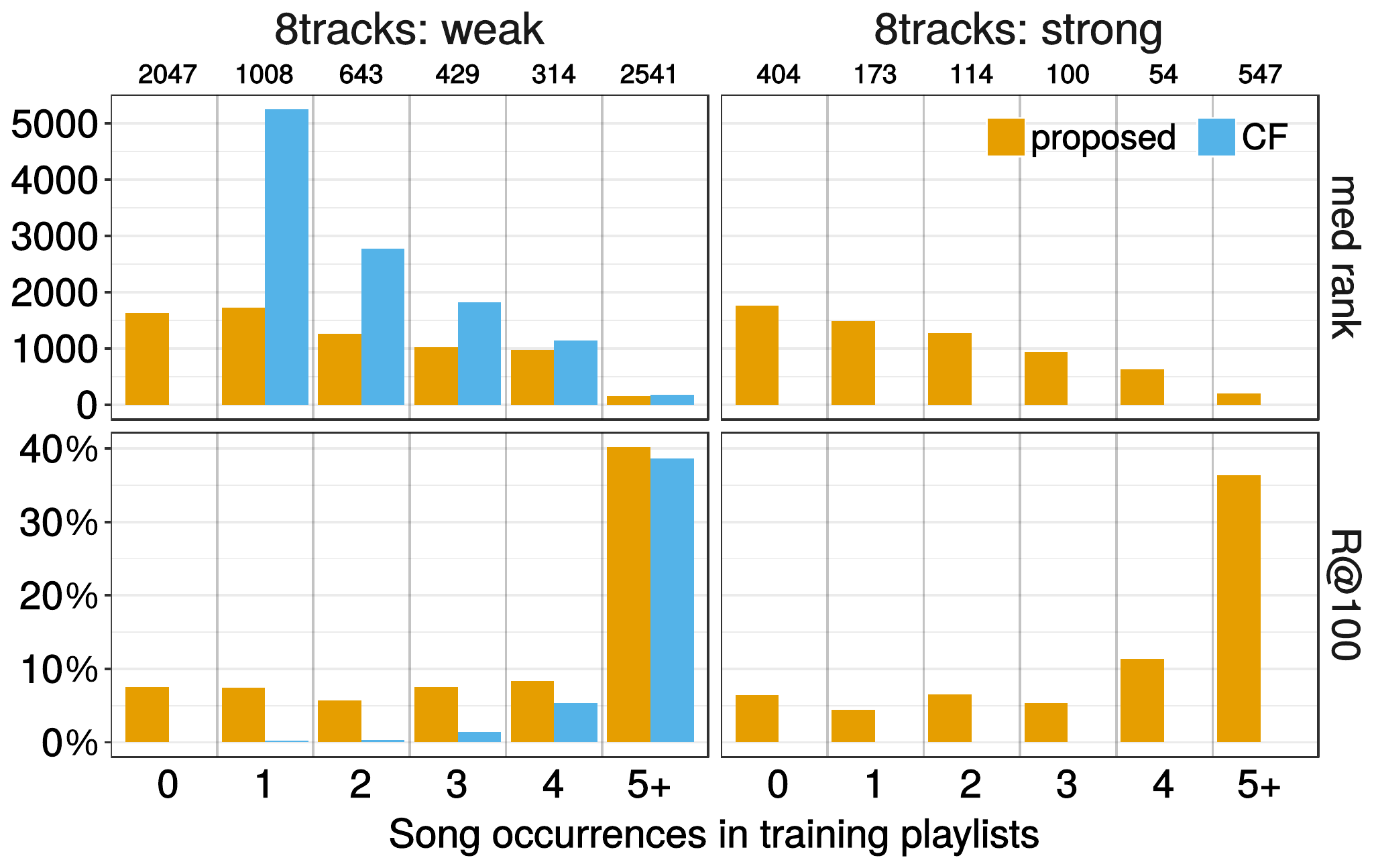}        
    \end{subfigure}
	\caption{\small Results achieved by the proposed playlist continuation model and the CF baseline on the off-line experiment as a function of how often the songs in the withheld continuations occurred in training playlists. We report the median rank (lower is better) and the recall at 100 (R@100, higher is better). The text annotations on top of each panel indicate the number of observations in each bar, which are the same for the proposed model and the CF baseline.}
    \label{sac2018_membership:figs:coldsongs}
\end{figure*}

We now analyze the robustness of the proposed playlist continuation model to recommend rare and out-of-set songs. Figure~\ref{sac2018_membership:figs:coldsongs} shows the results from the off-line experiments discussed above as a function of how often the songs in the withheld continuations occurred in training playlists. We first focus on the weak generalization results, where we can compare the proposed model and the CF baseline. As expected, both models achieve best performances when they recommend songs that occurred in 5 or more training playlists. As we know, the CF baseline can not recommend out-of-set songs, but we also observe that is has severe difficulties to recommend songs occurring rarely in training playlists. The performance of the proposed model on rare and out-of-set songs is not as good as on songs occurring in five or more playlists. Nevertheless, the proposed model is able to rank such songs much better than a random model would (Table~\ref{sac2018_membership:table:results}) and clearly better than the CF baseline.


We finally focus on the robustness of the proposed playlist continuation model to recommend rare and out-of-set songs in the strong generalization setting. As before, we want to investigate if the proposed model's performance is harmed in this evaluation setting, compared to its performance in the weak generalization setting. Still in Figure~\ref{sac2018_membership:figs:coldsongs}, we observe that the behavior of the proposed model in the weak and strong generalization settings is fairly similar, in line with the overall results discussed above~(Table~\ref{sac2018_membership:table:results}). This is an interesting finding because it provides an empirical indication that regarding playlist-song pairs exclusively in terms of their feature representations favors generalization and discourages the specialization towards particular training playlists.

\section{Conclusion}
\label{sac2018_membership:sec:conclusion}

We have introduced a novel hybrid playlist continuation model based on the general notion of playlist-song membership. The model integrates collaborative information encoded in curated playlists with state-of-the-art multimodal song features derived from audio, social tags and independent listening logs. By design, it regards playlists-song pairs exclusively in terms of their feature vectors, seeking to discourage the specialization towards specific playlists and songs. As a consequence, in contrast to CF models limited to extending previously profiled playlists with songs seen at training time, the proposed playlist continuation model can flexibly decide the fitness of any playlist-song pair, regardless of whether the playlist and the song were observed at training time. Furthermore, we follow a feature-combination hybrid approach, that is, the different sources of information are implicitly fused into a standalone enhanced playlist continuation model. According to our experimental results, the proposed model compares to a state-of-the art CF model for the task of extending profiled playlists, when sufficient training data is available. In contrast to the CF model, the proposed model can additionally extend non-profiled playlists without a significant performance loss, and recommend rare and out-of-set songs with fair performance. We believe that the current work provides a proof of concept for this newly proposed approach to playlist modeling. The natural next step is the comprehensive benchmarking of the proposed approach against other competing techniques and the identification of improvable aspects.


\section*{Acknowledgments}

We thank Christine Bauer, Hamid Eghbal-Zadeh and Roc\'{i}o del R\'{i}o Lorenzo for their valuable comments and the helpful discussions. This research has received funding from the European Research Council (ERC) under the European Union's Horizon 2020 research and innovation programme under grant agreement No 670035.

\bibliographystyle{ACM-Reference-Format}
\bibliography{sac2018_membership}

\end{document}